\documentclass[aps,prd,preprint,showpacs,groupedaddress]{revtex4}

\usepackage{graphicx}

\def\PL{\left(\frac{1-\gamma_5}{2}\right)}
\def\PR{\left(\frac{1+\gamma_5}{2}\right)}
\def\r2{\sqrt 2}
\def\tanW{\tan\theta_W}
\def\tanw#1{\tan^#1\theta_W}
\def\sinw#1{\sin^#1\theta_W}
\def\cosw#1{\cos^#1\theta_W}
\def\tanb{\tan\beta}
\def\cosb2{\cos^2\beta}
\def\Usn{\tilde U_\nu}
\def\Un{U_\nu}
\def\la{l_\alpha}
\def\lb{l_\beta}
\def\na{\nu_\alpha}

\def\w{\omega}
\def\x{\chi}
\def\sl{\tilde l}
\def\sn{\tilde \nu }
\def\m#1{{\tilde m}_#1}
\def\mla{m_{l_\alpha}}
\def\ml#1{m_{l_#1}}
\def\mn#1{m_{\nu_#1}}
\def\mwi{m_{\omega_i}}
\def\mw#1{m_{\omega_#1}}
\def\mwk{m_{\omega_k}}
\def\mwl{m_{\omega_l}}
\def\mx#1{m_{\chi_#1}}
\def\mxn{m_{\chi_n}}
\def\Msn#1{M_{\tilde\nu_#1}}
\def\Mn{M_{\tilde\nu}}
\def\Mna{M_{\tilde\nu_a}}
\def\Mnap{M_{\tilde\nu_{a'}}}
\def\Mnb{M_{\tilde\nu_b}}
\def\Mnbp{M_{\tilde\nu_{b'}}}
\def\Gn{\Gamma_{\tilde\nu}}
\def\Gna{\Gamma_{\tilde\nu_a}}
\def\Gnb{\Gamma_{\tilde\nu_b}}

\def\PRD#1#2#3{Phys. Rev. {\bf D#1}, #2 (#3)}
\def\NPB#1#2#3{Nucl. Phys. {\bf B#1}, #2 (#3)}

\def\EPJC#1#2#3{Eur. Phys. J. {\bf C#1}, #2 (#3)}
\def\PLB#1#2#3{Phys. Lett. {\bf B#1}, #2 (#3)}
\def\PRL#1#2#3{Phys. Rev. Lett. {\bf #1}, #2 (#3)}

\begin{document}

\preprint{
OCHA-PP-235
}

\title{
Generation-changing interaction of sneutrinos in $e^+e^-$ collisions  
}


\author{
Noriyuki Oshimo
}
\affiliation{
Institute of Humanities and Sciences {\rm and} Department of Physics \\
Ochanomizu University, Tokyo, 112-8610, Japan
}


\date{\today}

\begin{abstract}

     Measurability of generation mixing is studied on the pair production 
of sneutrinos in $e^+e^-$ collisions and their subsequent decays into two 
different charged leptons $e$ and $\mu$ with two lighter charginos.  
The analyses are made systematically in a general framework 
of the supersymmetric extension of the standard model.  
The production and decay process depends on the parameters of 
the chargino sector, as well as those of the sneutrino sector.   
Although generation-changing interactions are severely constrained by 
radiative charged-lepton decays, sizable regions in the parameter space 
could still be explored at $e^+e^-$ colliders in near future.  
 
\end{abstract}


\maketitle


\section{Introduction}

     Mixing of quarks or leptons belonging to different generations 
is suggestive.  
It could give a clue to underlying theories for the 
generations.  
For this end, we should get phenomenological information about the mixing 
as much as possible.   
Until recently most experimental data for the mixing were on the 
quarks, though data on the leptons are accumulating from the neutrino 
oscillations \cite{neutrino}.  
As a result, for instance, an examination of grand unified theories 
can now be made through comparison of the mixing for quarks and 
leptons \cite{oshimo}.  

     The supersymmetric standard model (SSM) is considered 
a plausible candidate for physics beyond the standard model (SM).  
This model includes superpartners of quarks and leptons, 
which could also be mixed among different generations.  
If the SSM is indeed the extension of the SM, 
information about generation mixing of these supersymmetric particles  
will also help us to have deep insight into the generations.   
In future experiments, therefore, study of generation 
mixing in the SSM may well be an important subject.  
It has been shown \cite{hisano} that a pair of different charged leptons,  
especially $\tau$ and $\mu$ or $e$, could be produced at detectable 
rates in $e^+e^-$ collisions, assuming the slepton generation mixing 
predicted by a certain scenario for the SSM.   

     In this article we study the generation-changing interaction of 
charged leptons, sneutrinos, and charginos without assuming a specific 
scenario for generation mixing.   
This interaction can create two different charged leptons with two 
charginos in $e^+e^-$ annihilation through the sneutrinos on mass-shell,   
which may be measured in near future experiments.  
We obtain the cross section of this process within the general 
framework of the SSM.   
The same interaction, on the other hand, induces radiative charged-lepton  
decays, which are constrained by non-observation in experiments 
to date \cite{pdg}.   
Under these constraints, concentrating on the production of $e$ and $\mu$ 
with two lighter charginos, systematic analyses are performed to discuss 
the possibility of measuring the interaction at $e^+e^-$ colliders.  
Particular attention is payed to the dependencies on various SSM parameters.   
It is shown that there are sizable regions of the parameter space 
which could be examined in the future experiments.   

     In $e^+e^-$ annihilation a pair of sneutrinos $\sn_a$ are created 
if kinematically allowed:  $e^+e^-\to\sn_a\sn_b^*$, 
where the indices $a$ and $b$ represent the generations.  
Assuming that the sneutrinos are heavier than some of the charginos $\w_i$,  
the sneutrino can decay into a charged lepton and a chargino 
$\sn_a^{(*)}\to\la^{-(+)}\w_i^{+(-)}$, 
with $\alpha$ representing the generation.  
The generation mixing could be measured by tagging the charged leptons.  
However, it is suggested by both theoretical considerations and 
experimental constraints derived from radiative charged lepton decays  
that some sneutrino masses are highly degenerated.   
Then, the generation mixing has to be analyzed without specifying 
the sneutrino generations \cite{arkani}.  
Our analyses are made generically under an assumption that the sneutrinos 
of three generations are produced at the same collision energy and   
all the sneutrinos can yield a charged lepton of any generation.    

     This paper is organized as the following.  
In Sect. \ref{sect2}, the interactions of sneutrinos relevant 
to their pair production and dominant two-body decays are summarized.   
In Sect. \ref{sect3} the cross section is given for the production 
of two charged leptons and two charginos mediated by real sneutrinos 
in $e^+e^-$ annihilation.   
In Sect. \ref{sect4}, numerical analyses for the production of $e$ and $\mu$ 
with two lighter charginos are performed, discussing measurability of 
generation mixing.  
Summary is given in Sect. \ref{sect5}.

\section{Interactions \label{sect2}}

     The mass eigenstates of the leptons or the sneutrinos are 
generally not the same as their interaction eigenstates.  
Assuming that no superfield for a right-handed neutrino exists 
at the electroweak energy scale, the sneutrinos have a $3\times 3$ 
hermitian mass-squared matrix $\tilde M_\nu^2$.   
The generation mixing of the sneutrinos is traced back to soft 
supersymmetry-breaking terms.  
The mass matrix $M_l$ for the charged leptons is 
proportional to the coefficient matrix of the Higgs couplings.   
For the left-handed neutrinos, we assume a Majorana mass matrix $M_\nu$.   
These mass matrices for the leptons are $3\times 3$ and non-diagonal.  
The mass eigenstates are obtained by diagonalizing these matrices as 
\begin{eqnarray}
\Usn^\dagger \tilde M_\nu^2\Usn &=& {\rm diag}(\Msn1^2, \Msn2^2, \Msn3^2),  \\ 
U_{lR}^\dagger M_lU_{lL} &=& {\rm diag}(\ml1, \ml2, \ml3),  \\
\Un^T M_\nu \Un &=& {\rm diag}(\mn1, \mn2, \mn3), 
\end{eqnarray}
where $\Usn$, $U_{lR}$, $U_{lL}$, and $\Un$ are unitary matrices.   
The masses of the charged leptons $e$, $\mu$, and $\tau$ are denoted by  
$\ml1$, $\ml2$, and $\ml3$, respectively. 
For the parameters which describe the sneutrino mass-squared matrix, we take 
the mass eigenvalues $\Mna$ $(a=1,2,3)$ and the unitary matrix 
$\tilde U_\nu$.  

     The charginos $\w_i$ $(i=1,2)$ and the neutralinos $\x_n$ $(n=1-4)$ 
are the mass eigenstates for the SU(2)$\times$U(1) gauginos and Higgsinos.  
The mass matrices $M^-$ and $M^0$ for the charginos and the neutralinos, 
respectively, are given by 
\begin{eqnarray}
    M^- &=& \left(
      \matrix{           \m2     & -\frac{1}{\r2}gv_1 \cr
               -\frac{1}{\r2}gv_2 & m_H                   }        
           \right),
\label{chmass}  \\ 
   M^0 &=& \left(
\matrix{     \m1  &               0  & \frac{1}{2}g'v_1 & -\frac{1}{2}g'v_2 \cr
               0  &              \m2 & -\frac{1}{2}gv_1 &  \frac{1}{2}gv_2  \cr
 \frac{1}{2}g'v_1 & -\frac{1}{2}gv_1 &               0  &         -m_H      \cr
-\frac{1}{2}g'v_2 &  \frac{1}{2}gv_2 &             -m_H &           0         }
           \right).
\label{nemass}
\end{eqnarray}
Here, $\m1$ and $\m2$ stand for the gaugino mass parameters for 
respectively U(1) and SU(2), for which we assume the relation 
$\m1=(5/3)\tanw2\m2$ suggested by the SU(5) grand unified theory.  
The Higgsino mass parameter is denoted by $m_H$.  
The vacuum expectation values of the Higgs bosons with hypercharges 
$-1/2$ and $1/2$ are expressed by $v_1$ and $v_2$, respectively, 
the ratio $v_2/v_1$ being denoted by $\tanb$.   
These mass matrices are diagonalized to give mass eigenstates as
\begin{eqnarray}
      C_R^\dagger M^-C_L &=& {\rm diag}(\mw1, \mw2), \\
                       & & \mw1 <\mw2,   \nonumber \\
N^TM^0N &=& {\rm diag}(\mx1, \mx2, \mx3, \mx4), \\
                       & & \mx1<\mx2<\mx3<\mx4,  \nonumber 
\end{eqnarray}
where $C_R$, $C_L$, and $N$ are unitary matrices.

     We express the interaction Lagrangians for the sneutrinos $\sn_a$ 
in terms of particle mass eigenstates.  
The Lagrangian for charged leptons, sneutrinos, and charginos is given by 
\begin{eqnarray}
\cal L &=& i\frac{g}{\r2}(V_C)_{a\alpha}\sn_a^\dagger\overline{\w_i}
\left[\r2 C_{R1i}^*\PL+C_{L2i}^*\frac{m_{l_\alpha}}{\cos\beta M_W}\PR\right] 
l_\alpha  +{\rm H.c.},  
\label{cint}
\end{eqnarray}
with $V_C=\Usn^\dagger U_{lL}$.  
The generation mixing is described by $V_C$, which is a $3\times 3$ 
unitary matrix.  
For the independent physical parameters of $V_C$ three mixing angles 
and one complex phase can be taken, the other complex phases being 
left out by redefinition of particle fields.   
We adopt the parametrization in the standard form for the 
Cabibbo-Kobayashi-Maskawa matrix: 
\begin{equation}
V_C = \left(\matrix{c_{12}c_{13} & s_{12}c_{13} &
                    s_{13}e^{-i\delta}  \cr
      -s_{12}c_{23}-c_{12}s_{23}s_{13}e^{i\delta} &
      c_{12}c_{23}-s_{12}s_{23}s_{13}e^{i\delta} &
          s_{23}c_{13}             \cr
      s_{12}s_{23}-c_{12}c_{23}s_{13}e^{i\delta} &
      -c_{12}s_{23}-s_{12}c_{23}s_{13}e^{i\delta} &
          c_{23}c_{13}       }    \right),
\end{equation}
with $c_{ab}=\cos\theta_{ab}$ and $s_{ab}=\sin\theta_{ab}$.
Without loss of generality, the angles $\theta_{12}$, $\theta_{23}$, 
and $\theta_{13}$ can be put in the first quadrant. 
The Lagrangian for neutrinos, sneutrinos, and neutralinos is given by 
\begin{eqnarray}
\cal L &=& i\frac{g}{\r2}(V_N)_{a\alpha}(-\tanW N_{1n}+N_{2n})
\sn_a^\dagger\overline{\x_n}\PL \nu_\alpha  +{\rm H.c},  
\label{nint}
\end{eqnarray}
with $V_N=\Usn^\dagger \Un$.  
The $3\times 3$ unitary matrix $V_N$ describes the generation mixing.   
The number of its physical parameters is six, though a definite 
parametrization is not necessary for our analyses.  
Finally the coupling with the $Z$ boson is expressed by the Lagrangian
\begin{eqnarray}
\cal L &=& -i\frac{\sqrt{g^2+g'^2}}{2}
       \left(\sn_a^*\partial^\mu\sn_a-\partial^\mu\sn_a^*\sn_a\right)Z_\mu, 
\end{eqnarray}
which does not cause generation-changing interaction.  

\section{Cross section \label{sect3}}

     A pair of sneutrinos are created in $e^+e^-$ annihilation 
by the $Z$-boson and chargino exchange diagrams.  
We assume that the masses of the sneutrinos are 
almost degenerated among three generations.  
The electron and positron couple to all the sneutrinos through 
charginos as shown in Eq. (\ref{cint}).  
Therefore, a pair of sneutrinos in any combination of generations 
can be produced at a collision energy above the threshold.   
The sneutrino dominantly decays into a charged lepton and a chargino 
or into a neutrino and a neutralino.   
These leptons can also belong to any generation, owing to the 
interactions in Eqs. (\ref{cint}) and (\ref{nint}).   
   
     We give the cross section for $e^+e^-\to \la^-\lb^+\w^+_i\w^-_j$ 
mediated by the sneutrinos on mass-shell.  
In calculating this cross section the products of the sneutrino 
propagators are involved.  
We make an approximation:   
\begin{eqnarray}
  \frac{1}{(q^2-\Mna^2+i\Mna\Gna)(q^2-\Mnb^2-i\Mnb\Gnb)} &=&  
 \frac{\pi\left\{\delta(q^2-\Mna^2)+\delta(q^2-\Mnb^2)\right\}}
   {2(1+ix_{ab})(\overline{\Mn\Gn})_{ab}},  \\
 (\overline{\Mn\Gn})_{ab} &=& \frac{\Mna\Gna+\Mnb\Gnb}{2},  \nonumber \\ 
  x_{ab} &=& \frac{\Mna^2-\Mnb^2}{\Mna\Gna+\Mnb\Gnb}, \nonumber 
\end{eqnarray}
where $\Gna$ denotes the total decay width of the sneutrino $\sn_a$.   
The relations $\Mna\gg\Gna$, $\Mnb\gg\Gnb$, and 
$|\Mna^2-\Mnb^2|\ll\Mna^2\approx\Mnb^2$ have been assumed.  
The cross section is written as 
\begin{eqnarray}
 & & \sigma(e^+e^-\to \la^-\lb^+\w^+_i\w^-_j) = \sum_{a,a',b,b'}  
 \frac{(V_C^\dagger)_{\alpha a}(V_C)_{a'\alpha}
                  (V_C)_{b\beta}(V_C^\dagger)_{\beta b'}}
{4(1+ix_{aa'})(1+ix_{bb'})(\overline{\Mn\Gn})_{aa'}(\overline{\Mn\Gn})_{bb'}}  
                         \nonumber \\
 & & \Bigl\{\tilde\sigma_{aa'bb'}(\Mna,\Mnb)\Mna\Mnb
\tilde\Gamma(\sn_a\to \la\w_i)\tilde\Gamma(\sn_b\to \lb\w_j) 
                  \nonumber \\
 & & + \tilde\sigma_{aa'bb'}(\Mna,\Mnbp)\Mna\Mnbp
\tilde\Gamma(\sn_a\to \la\w_i)\tilde\Gamma(\sn_{b'}\to \lb\w_j) 
                    \nonumber \\
 & & + \tilde\sigma_{aa'bb'}(\Mnap,\Mnb)\Mnap\Mnb
\tilde\Gamma(\sn_{a'}\to \la\w_i)\tilde\Gamma(\sn_b\to \lb\w_j) 
                       \nonumber \\
 & & + \tilde\sigma_{aa'bb'}(\Mnap,\Mnbp)\Mnap\Mnbp
\tilde\Gamma(\sn_{a'}\to \la\w_i)\tilde\Gamma(\sn_{b'}\to \lb\w_j)\Bigr\}, 
\end{eqnarray}
where the summation for each index is done over three generations.  
Expressing the total energy at the center-of-mass frame by $\sqrt{s}$, 
the factor $\tilde\sigma_{aa'bb'}(\Mna,\Mnb)$ is defined by   
\begin{eqnarray}
  && \tilde\sigma_{aa'bb'}(\Mna,\Mnb) =   
   \frac{g^4}{64\pi s^2}\int_{t_-}^{t_+}dt\left(tu-\Mna^2\Mnb^2\right)  
                     \nonumber \\ 
  & & \Bigl[\delta_{ab}\delta_{a'b'}
\left\{\left(\frac{-1+2\sinw2}{2\cosw2}\right)^2+\tanw4\right\}
        \frac{1}{(s-M_Z^2)^2}   
                   \nonumber  \\
  &+&(V_C)_{a1}(V_C^\dagger)_{1b}(V_C^\dagger)_{1a'}(V_C)_{b'1}
    \sum_{k,l}\frac{|C_{R1k}|^2|C_{R1l}|^2}{(t-\mwk^2)(t-\mwl^2)}  
                    \nonumber \\ 
 &+&\left\{\delta_{ab}(V_C^\dagger)_{1a'}(V_C)_{b'1}
  +(V_C)_{a1}(V_C^\dagger)_{1b}\delta_{a'b'}\right\}
                \frac{-1+2\sinw2}{2\cosw2}
      \sum_k\frac{|C_{R1k}|^2}{(t-\mwk^2)(s-M_Z^2)}  \Bigr], 
                      \nonumber  \\
 u &=& -s-t+\Mna^2+\Mnb^2,   \\
 t_\pm &=& \frac{1}{2}\left\{\Mna^2+\Mnb^2-s\pm
    \sqrt{s^2-2(\Mna^2+\Mnb^2)s+(\Mna^2-\Mnb^2)^2}\right\},  
       \nonumber
\end{eqnarray}
which arises from the sneutrino pair production.  
The sneutrino decay yields the factor  
$\tilde\Gamma(\sn_a\to \la\w_i)$, which is defined by   
\begin{eqnarray}
  \tilde\Gamma(\sn_a\to \la\w_i) 
 &=& \frac{g^2}{16\pi}\Mna\sqrt{
      \left\{1-\frac{(\mla+\mwi)^2}{\Mna^2}\right\}
      \left\{1-\frac{(\mla-\mwi)^2}{\Mna^2}\right\}}  
                  \nonumber   \\
 & & \Bigl\{\left(|C_{R1i}|^2+|C_{L2i}|^2\frac{\mla^2}{2\cosb2 M_W^2}\right)
                \left(1-\frac{\mla^2+\mwi^2}{\Mna^2}\right)
               \nonumber \\
 & & -{\rm Re}\left[C_{R1i}C_{L2i}^*\right]\frac{2\r2\mla^2\mwi}
                {\cos\beta M_W\Mna^2}\Bigr\}.   
\end{eqnarray}
The total width of the sneutrino is approximately 
determined by the two-body decays, 
\begin{eqnarray}
  \Gna &=& \sum_{\alpha,i}\Gamma(\sn_a\to \la^-\w^+_i) +  
             \sum_{\alpha,n}\Gamma(\sn_a\to \na\x_n),  \\ 
   & &  \Gamma(\sn_a\to \la^-\w^+_i) = |(V_C)_{a\alpha}|^2
                           \tilde\Gamma(\sn_a\to \la\w_i),  
                \nonumber \\ 
   & &  \Gamma(\sn_a\to \na\x_n) = |(V_N)_{a\alpha}|^2
           \frac{g^2}{32\pi}\Mna\left|-\tanW N_{1n}+N_{2n}\right|^2
             \left(1-\frac{\mxn^2}{\Mna^2}\right)^2,  
                  \nonumber 
\end{eqnarray}
where the masses of the neutrinos have been neglected.  
Since the equality $\sum_\alpha|(V_N)_{a\alpha}|^2=1$ holds, 
the mixing matrix $V_N$ need not to be specified for obtaining $\Gna$.  

     The radiative charged lepton decays $\mu\to e\gamma$, 
$\tau\to e\gamma$, and $\tau\to\mu\gamma$ are induced at one-loop 
level by the exchange of charginos and sneutrinos.     
For calculation of these decay widths, we refer to Appendix of 
Ref. \cite{bsg}.  
Applying the given formulae to the interaction in Eq. (\ref{cint}), 
the decay widths are obtained straightforwardly.  
The radiative decays are also generated by one-loop diagrams 
with charged sleptons and neutralinos.  
However, these contributions are generally smaller than those of 
chargino-sneutrino diagrams, so that we neglect the former ones.   

\section{Numerical analyses \label{sect4}}

     Specializing in the process 
$e^+e^-\to\sum_{a,b}\sn_a\sn_b^*\to e^-\mu^+\w^+_1\w^-_1$,  
we discuss numerically its cross section.  
The intermediate sneutrinos are on mass-shell, belonging to 
any generations.  
This process shows distinctive final states.  
The primary leptons $e$ and $\mu$ are produced by two-body decays.   
Their energies are large, unless the mass of the lighter chargino is 
close to the sneutrino masses.  
In addition, they have approximately the same value and are monochromatic 
in each rest frame of the decaying sneutrino.    
The chargino yields, by three-body decays, two quarks or two leptons 
with a neutralino.  
Therefore, the final states involve, in each hemisphere, one charged lepton 
with a large energy of flat distribution and two jets or one charged lepton, 
together with large missing energy-momentum.   
The detection of the process will not be difficult.   
The collider energy is assumed as $\sqrt{s}=500$ GeV throughout 
the analyses.  

\begin{figure}
\includegraphics{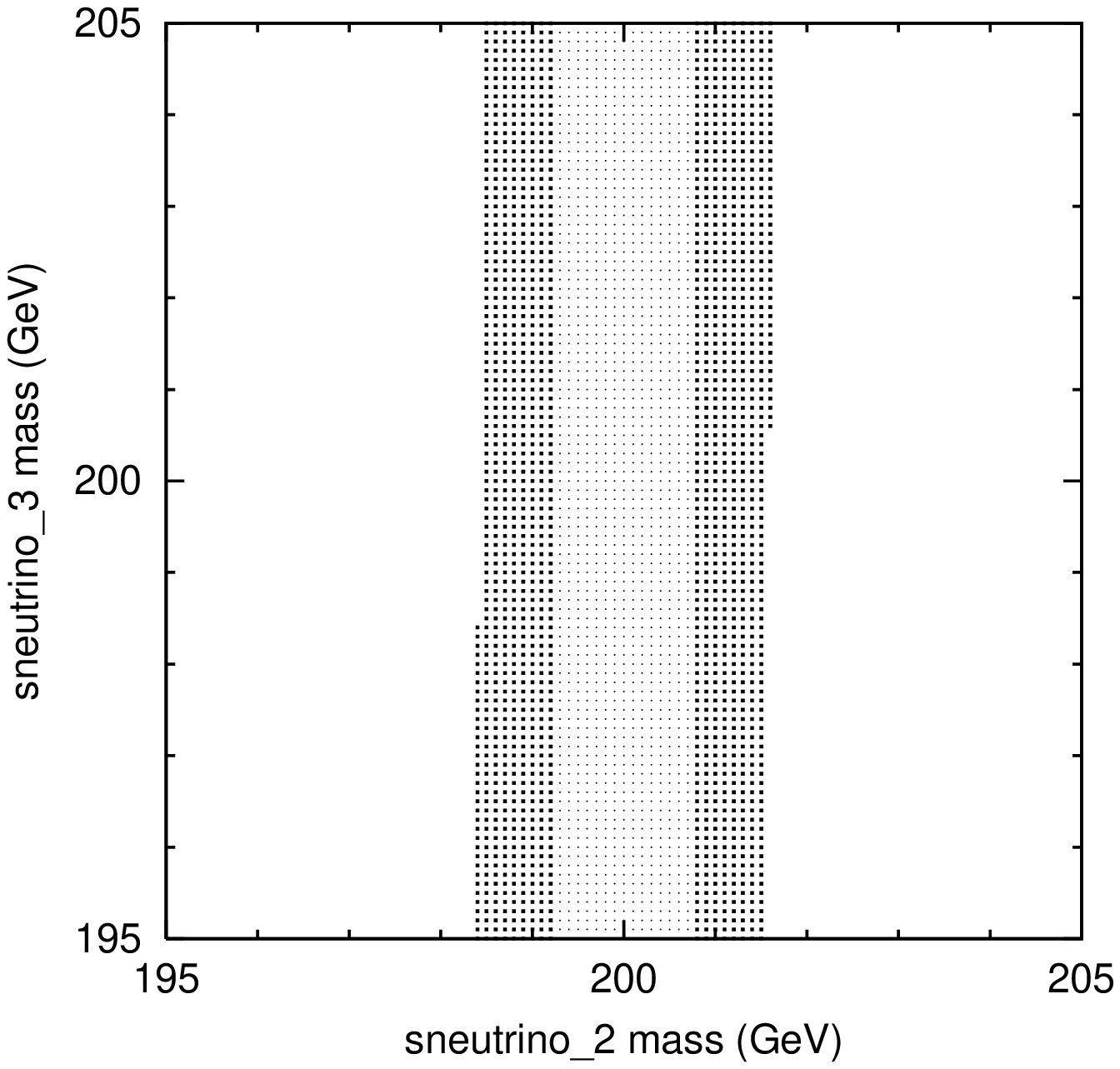}%
\caption{
The cross section for the parameter set (a) listed in Table \ref{table}:   
$\sigma<$ 0.05 fb in the light shaded region and 
0.05 fb $<\sigma<$ 0.07 fb in the dark-shaded region. 
$\Msn1=200$ GeV, $\theta_{12}=\theta_{23}=\theta_{13}=0.02$, 
and $\delta=\pi/4$.
\label{emu1}
   }
\end{figure}

\begin{figure}
\includegraphics{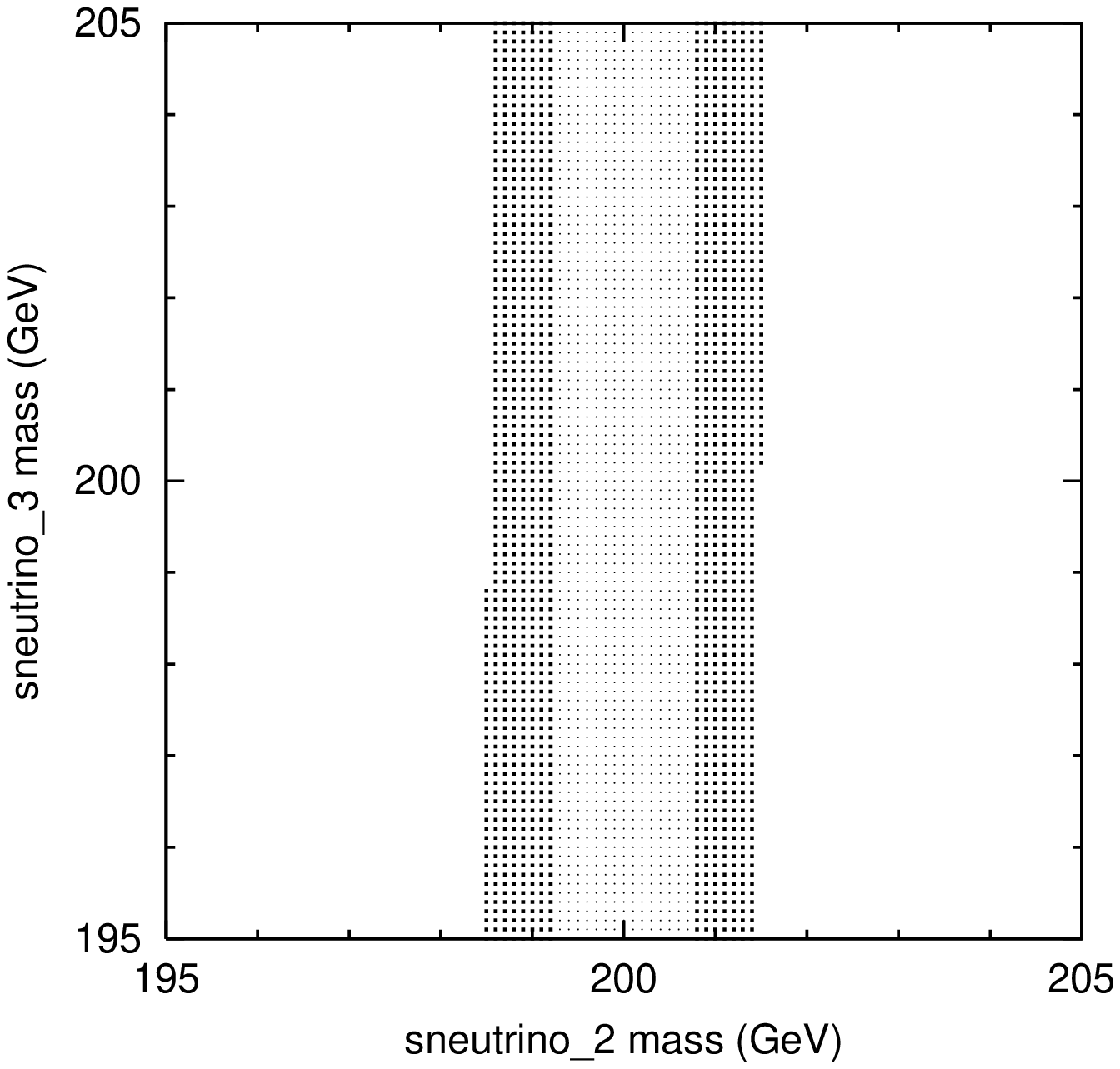}%
\caption{
The cross section for the parameter set (b) listed in Table \ref{table}:   
$\sigma<$ 0.05 fb in the light shaded region and 
0.05 fb $<\sigma<$ 0.1 fb in the dark-shaded region. 
$\Msn1=200$ GeV, $\theta_{12}=\theta_{23}=\theta_{13}=0.02$, 
and $\delta=\pi/4$.
\label{emu2}
   }
\end{figure}

\begin{figure}
\includegraphics{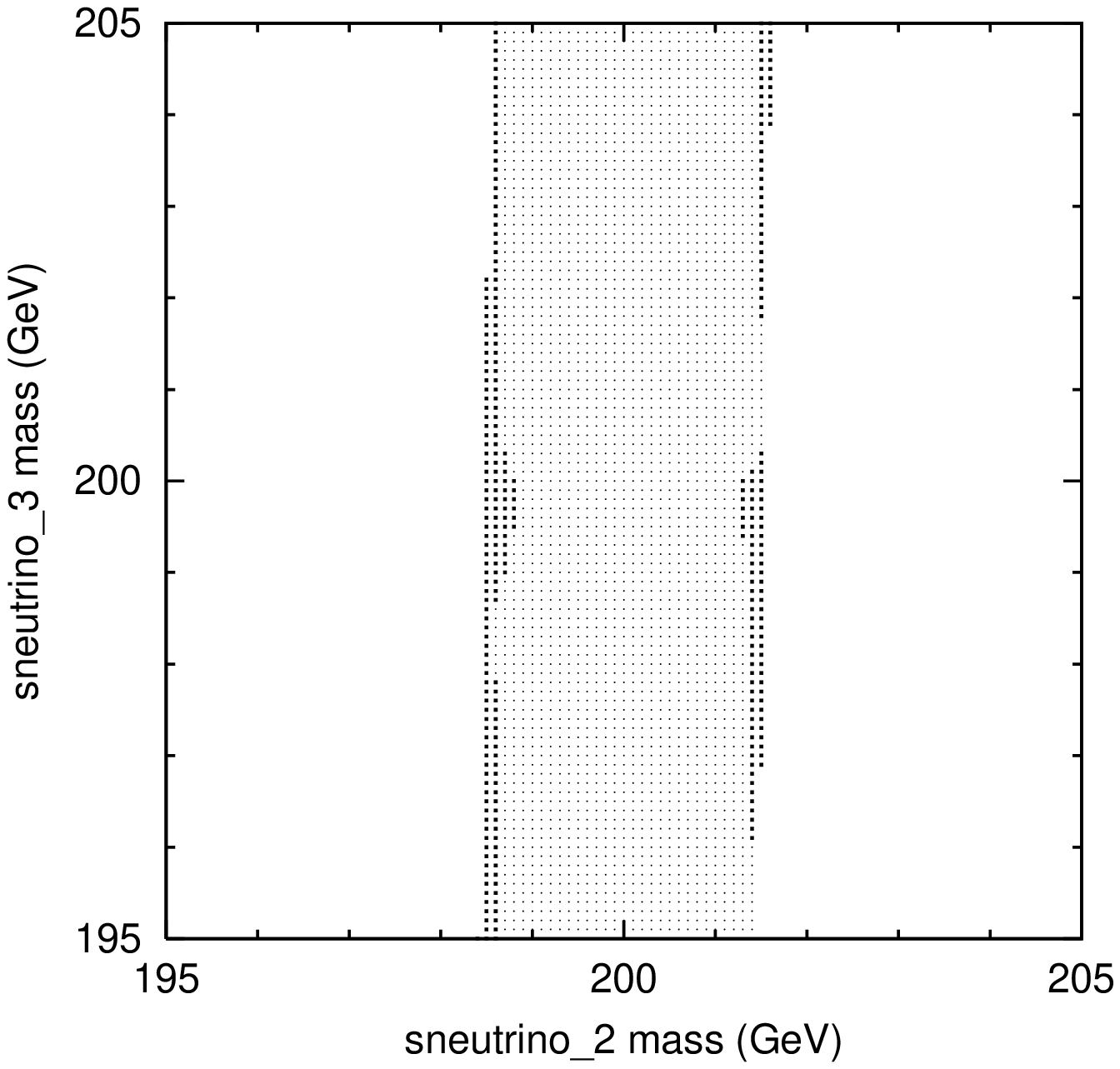}%
\caption{
The cross section for the parameter set (c) listed in Table \ref{table}:   
$\sigma<$ 0.05 fb in the light shaded region and 
0.05 fb $<\sigma<0.06$ fb in the dark-shaded region. 
$\Msn1=200$ GeV, $\theta_{12}=\theta_{23}=\theta_{13}=0.02$, 
and $\delta=\pi/4$.
\label{emu3}
   }
\end{figure}

     The cross section in discussion depends on mass differences 
of the sneutrinos, as the radiative charged-lepton decays do so.  
In Figs. \ref{emu1}, \ref{emu2}, and \ref{emu3} we show, in the 
$\Msn2$-$\Msn3$ plane, the cross section for the regions 
allowed by the constraints from the radiative charged-lepton decays.  
The mass of $\sn_1$ is fixed at $\Msn1=200$ GeV.  
The mixing angles and complex phase of $V_C$ are put at  
$\theta_{12}=\theta_{23}=\theta_{13}=0.02$ and $\delta=\pi/4$.  
The parameter values for the charginos and neutralinos are 
listed in Table \ref{table}, the sets (a), (b), and (c) corresponding 
to the figures \ref{emu1}, \ref{emu2}, and \ref{emu3}, respectively.   
The resultant masses of the charginos and neutralinos are also 
given in the table.    
The cross section has a value $\sigma<$ 0.05 fb in the light 
shaded regions, and in the dark shaded regions 
0.05 fb $<\sigma<$ 0.07 fb, 0.05 fb $<\sigma<$ 0.1 fb, and  
0.05 fb $<\sigma<$ 0.06 fb for Figs. \ref{emu1}, \ref{emu2}, 
and \ref{emu3}, respectively.  
Unshaded regions are excluded by the radiative 
charged-lepton decays.  

     As the mass difference between $\sn_1$ and $\sn_2$ becomes large, 
the cross section increases.  
The decay width of $\mu\to e\gamma$ also increases and 
becomes too large in the outside of the shaded region.  
The mass of $\sn_3$ does not sensitively affect the cross section nor 
the decay width.  
The constraints from the decays $\tau\to e\gamma$ and $\tau\to\mu\gamma$ 
are not very stringent, so that large mass differences between 
$\sn_1$ and $\sn_3$ and between $\sn_2$ and $\sn_3$ are allowed.  
Comparing three parameter choices for $\m2$ and $m_H$ of the sets (a), (b), 
and (c), we can see that the mass ranges of $\sn_2$ allowed by the 
radiative charged-lepton decays are not much different from each other.  
However, the cross section varies manifestly 
with these parameters, depending on the relative magnitudes of $\m2$ 
and $m_H$.  
For a smaller magnitude of $\m2/m_H$, the SU(2)-gaugino component of 
the lighter chargino is larger.  
Then, the cross section of $e^+e^-\to\sn_a\sn_b^*$ 
and thus that of $e^+e^-\to e^-\mu^+\w^+_1\w^-_1$ increase.  

\begin{figure}
\includegraphics{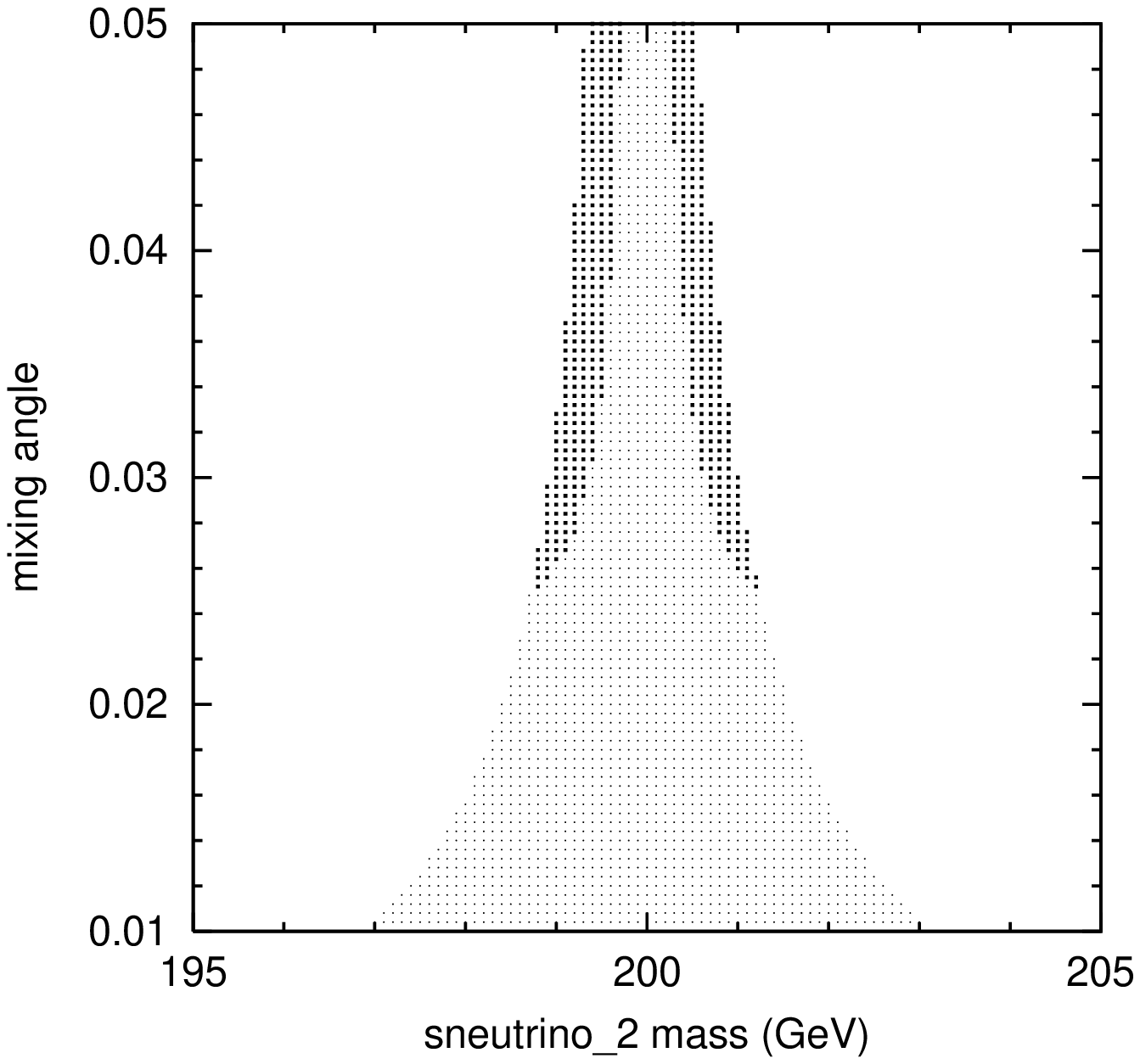}%
\caption{
The cross section for the parameter set (a) listed in Table \ref{table}:   
$\sigma<$ 0.1 fb in the light shaded region and 
0.1 fb $<\sigma<$ 0.3 fb in the dark-shaded region. 
$\Msn1=200$ GeV, $\Msn3=198$ GeV, and $\delta=\pi/4$.
\label{emu4}
   }
\end{figure}

\begin{figure}
\includegraphics{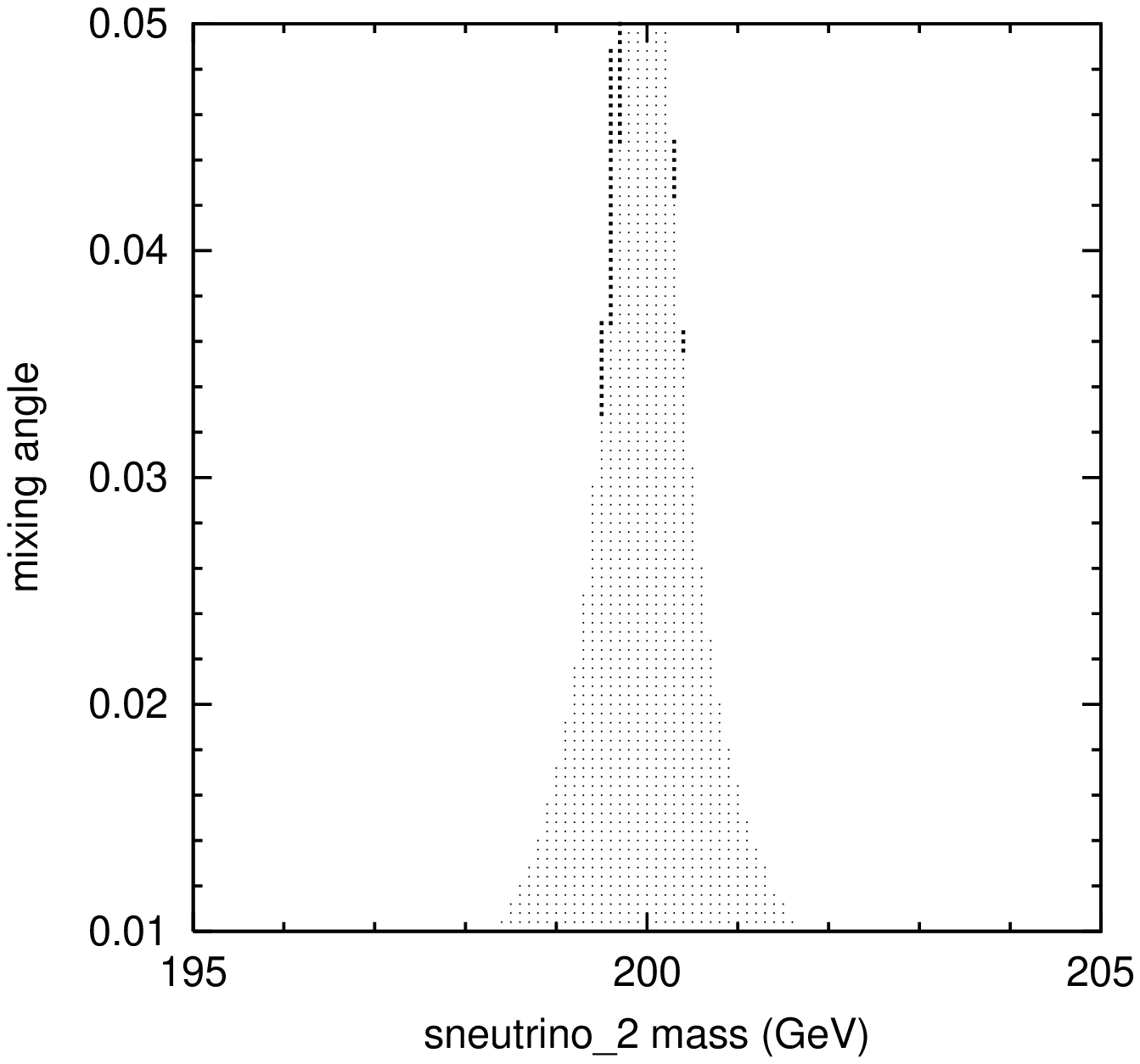}%
\caption{
The cross section for the parameter set (d) listed in Table \ref{table}:   
$\sigma<$ 0.1 fb in the light shaded region and 
0.1 fb $<\sigma<$ 0.2 fb in the dark-shaded region. 
$\Msn1=200$ GeV, $\Msn3=198$ GeV, and $\delta=\pi/4$.  
\label{emu5}
   }
\end{figure}

     The dependencies of the cross section on the mixing angles of 
$V_C$ and $\tanb$ are shown in Figs. \ref{emu4} and \ref{emu5}.  
Assuming the equality $\theta_{12}=\theta_{23}=\theta_{13}$ 
$(\equiv\theta)$ with $\delta=\pi/4$ for simplicity, the cross section 
is shown in the $\Msn2$-$\theta$ 
plane for the regions consistent with the radiative charged-lepton decays.  
The masses of $\sn_1$ and $\sn_3$ are fixed at $\Msn1=200$ GeV 
and $\Msn3=198$ GeV.  
The parameter values for the charginos and neutralinos are given 
by the sets (a) and (d) in Table \ref{table} for the figures 
\ref{emu4} and \ref{emu5}, respectively.   
The cross section has a value $\sigma<$ 0.1 fb in the light shaded regions, 
and in the dark shaded regions 0.1 fb $<\sigma<$ 0.3 fb and 
0.1 fb $<\sigma<$ 0.2 fb for Figs. \ref{emu4} and \ref{emu5}, respectively.  
As the mixing angle increases, the allowed range for the mass difference 
between $\sn_1$ and $\sn_2$ becomes narrow, 
while the cross section becomes large.  
These angle dependencies are primarily determined by $\theta_{12}$.  
Different values for $\theta_{23}$ and $\theta_{13}$ do not alter much 
the cross section and the allowed mass difference, as long as these 
mixing angles are small.  
For a larger value of $\tanb$, the allowed region becomes small,  
though the cross section does not vary much with it.  

     Under the constraints from the presently available experiments, 
the cross section of $e^+e^-\to e^-\mu^+\w^+_1\w^-_1$ is larger 
than 0.05 fb in sizable regions of the parameter space.  
It should be noted that there also exists a charge conjugate process 
which has the same cross section.   
For the integrated luminosity of 100 fb$^{-1}$, more than ten events 
are expected there.   
Although possible backgrounds have to be taken into account for 
estimating realistically available events, a number of order of
ten would not be insufficient for detection in near future experiments.    
 
     We comment on possible background processes. 
The two leptons $e$ and $\mu$ with missing energy-momentum can be produced 
by the $\tau^+\tau^-$ pair production and their subsequent leptonic 
decays, which is due to generation-conserving interaction of the SM.     
This pair production occurs as e.g. $e^+e^-\to ZW^+W^-$, 
$Z\to\tau^+\tau^-$ in the SM and 
$e^+e^-\to\sn_3\sn_3^*\to\tau^-\tau^+\w_1^+\w_1^-$ in the SSM.  
The final states of these processes are similar to those of the  
discussed generation-changing process.  
However, there are wide differences in magnitude and distribution of the 
produced charged-lepton energy between the decays of $\tau$ and $\sn_a$.  
An appropriate energy cuts will be useful to reduce the backgrounds.   
Next, suppose that the masses of the charged sleptons are not much 
different from the sneutrino masses and, in addition, larger than 
the mass of the second lightest neutralino.    
Then, a pair of charged sleptons are produced and can decay into 
two charged leptons with two second lightest neutralinos, 
$e^+e^-\to\sl_a\sl_b^*\to\la^-\lb^+\x_2\x_2$.  
Owing to possible generation-changing interaction of charged leptons, 
charged sleptons, and neutralinos, these two leptons could belong to 
different generations, such as $e$ and $\mu$.   
If the second lightest neutralino decays into two quarks with the 
lightest neutralino, the final state topology is the same as 
that of the lighter chargino decay which consists of two jets with 
missing energy-momentum.   
On the other hand, the second lightest neutralino does not lead to 
one charged lepton with missing energy-momentum.   
The background from the pair production of the charged sleptons is 
discarded by requiring that one additional charged lepton 
be contained in at least one hemisphere.  

     The production $e^+e^-\to e^-\mu^+\w^+_1\w^-_1$ could also be 
induced by the generation-changing interaction in Eq. (\ref{cint}) 
through the pair production of different charginos $\w_1$ and $\w_2$.   
The heavier chargino decays into one charged lepton and a sneutrino, and 
this sneutrino decays into another charged lepton and a lighter chargino, 
$\w_2^{+(-)}\to\la^{+(-)}\sn_a^(*)$, $\sn_a^(*)\to\lb^{-(+)}\w_1^{+(-)}$.  
This process yields $e$ and $\mu$ in one hemisphere and thus will be   
distinguishable from the previous process which leads to one charged 
lepton in each hemisphere.   
Therefore, the interaction could be examined independently by 
these two processes.  

\section{Summary \label{sect5}}

     We have discussed the possibility of measuring the generation mixing 
for the interaction of charged leptons, sneutrinos, and charginos 
in future $e^+e^-$ collision experiments.  
Numerical analyses have been made on the process 
$e^+e^-\to e^-\mu^+\w^+_1\w^-_1$.  
The experimental signal is given by a energetic charged lepton and two 
jets or a charged lepton with missing energy-momentum in each hemisphere.  
The experimental bound on the radiative decay $\mu\to e\gamma$ 
implies that the sneutrino mass difference 
and the mixing angle for the first two generations should be small.  
Under these constraints, the cross section of the process is 
larger than 0.05 fb in sizable regions of the parameter space.  
The SSM parameters for the chargino and neutralino sector affect 
non-trivially the cross section.  
Combined with information on charginos and neutralinos, 
the examination of the process will enable us  
to explore the generation mixing for the interaction.     
Or on the contrary, the study of the generation mixing will give 
information on the SSM parameters.   

\begin{acknowledgments}
     This work is supported in part by the Grant-in-Aid for 
Scientific Research on Priority Areas (No. 16028204) from the 
Ministry of Education, Science and Culture, Japan.   
\end{acknowledgments}



\newpage
\pagebreak


%

\begin{table}
\caption{
The parameter values for the charginos and neutralinos 
assumed in Figs. \ref{emu1}--\ref{emu5}.  
The unit of mass is GeV.  
\label{table}
 }
\begin{ruledtabular}
\begin{tabular}{ccccc}
  & $(a)$ & $(b)$ & $(c)$ & $(d)$   \\ 
\hline  
 $\tanb$ & 5 & 5 & 5 & 10 \\
 $\m2$  & 200 & 150 & 250 & 200  \\
 $m_H$  & 200 & 250 & 150 & 200  \\
\hline
 $\mw1$ & 138 & 121 & 121 & 144 \\ 
 $\mw2$ & 272 & 288 & 288 & 269 \\ 
\hline
 $\mx1$ & 83 & 65 & 88 & 86 \\
 $\mx2$ & 146 & 124 & 149 & 149 \\ 
 $\mx3$ & 207 & 257 & 157 & 209 \\
 $\mx4$ & 273 & 289 & 289 & 269 \\
\end{tabular}
\end{ruledtabular}
\end{table}


\newpage
\pagebreak

%

\end{document}